\documentclass[]{aa}

\usepackage{epsfig}
\usepackage[T1]{fontenc}
\usepackage{textcomp}
\usepackage[latin1]{inputenc}
\usepackage{amsfonts}
\usepackage{natbib}

\usepackage{color}
\usepackage{graphicx}
\usepackage{psfrag}

\def\bff#1{{\bf #1}}
\newcommand{\refe}{}
\newcommand{\refee}{}
\newcommand{\refeee}{\bff}

\def\GRS{GRS $1915$+$105$\ }
\def\XTE{XTE J$1550$-$564$\ }

\newcommand{\ie}{{\it i.e.}~}

\begin{document}

\title{\refe{A microquasar classification from a disk instability perspective}}

\author{P. Varniere\inst{1}
\and M. Tagger\inst{2}
\and J Rodriguez\inst{3}} 

\institute{AstroParticule \& Cosmologie (APC), UMR 7164, 
Universit\'e Paris Diderot, 10 rue Alice Domon et Leonie Duquet, 75205 Paris Cedex 13, France.
varniere@apc.univ-paris7.fr
\and Laboratoire de Physique et Chimie de l'Environnement et de l'Espace, Universit\' e d'Orl\' eans/CNRS, France
\and Laboratoire AIM, CEA/IRFU-CNRS/INSU-Universite Paris Diderot, 
CEA DSM/IRFU/SAp, Centre de Saclay,  F-91191 Gif-sur-Yvette, France.}

\date{Received <date> /
Accepted <date>}
 
\abstract{}
{The spectacular variability of microquasars has led to a long string of efforts in order to classify their observed behaviors  \refe{in a few states}. 
The progress made in the understanding
of the Quasi-Periodic Oscillations observed in these objects now makes it possible  to develop a new \refe{way to find order in their behavior}, based on the 
\refe{theorized} physical processes associated with these oscillations.  This will also have the interest of reuniting microquasars 
in a single classification \refe{based on the physical processes at work and therefore} independent of their specificities
 (mass, variation timescale, outburst history, etc.).  
 \refe{This classification is aimed to be a tool to further our understanding of microquasars behavior and not to replace phenomenological states.}}
{We start by considering three instabilities that can cause accretion in the disk. We compare the conditions for their development, and the Quasi-Periodic Oscillations they can be expected to produce, with the spectral states in which these Quasi-Periodic Oscillations are observed and sometimes coexist.}
{\refe{From the three instabilities that we proposed to explain the three states of \GRS\ we actually found the theoretical existence of {\bf four} states.
We compared those four states with observations and also how those four states can be seen in a model-independent fashion. Those four state can be used
to find an order in microquasar observations, based on the properties of the Quasi-Periodic Oscillations and the physics of the associated 
instabilities.} }

\keywords{ X-rays: binaries,   stars: individual (GRS $1915$$+$$105$, XTE J$1550$$-$$564$), accretion disks} 
\titlerunning{Instability Based Classification}
\maketitle

\section{Introduction}  
\label{sect:intro}
  Since the first observations of microquasars, their strong variability has led to successive attempts at classifying them in universal 
  spectral states, defined by observables such as the luminosity, the energy spectrum of the emission, and also the presence of 
  Quasi-Periodic Oscillations (QPOs).
Historically, the classification has involved five different  states, based on the shape of the  energy spectrum 
and the flux level in the $1$-$10$ keV band,  \citep[see for example][]{VdK94, N95}. The first classification attempts started 
with the observation of Cyg X-1 and
GX 339-4, and interpretations were based on the mass accretion rate of the black hole. As  $\dot{M}$
increased the source was thought to go from quiescent to low/hard, intermediate, high/soft and very high state.\\
These states can be described as follows:
\begin{description}    
\item[\tt Low/hard state] The spectrum is a power-law with a  photon index  of the order of $1.5$ to $1.9$    and an exponential cutoff around $100$ keV. The X-ray luminosity is estimated to be below $10$\% of the Eddington luminosity \citep{N95}.
\item[\tt High/soft state] In this state the total luminosity is higher; the spectrum is dominated by a soft, blackbody-like component with a characteristic temperature of the order    of $1$ keV. A power-law tail is also present but much less luminous   than in the hard state, and its photon index is close to $2.5$
\item[\tt Intermediate state]  This state is seen during transitions between the low/hard and high/soft states, with spectral characteristics ranging \lq in-between\rq~those of these states.
\item[\tt Very high state]  In some systems a state with very high luminosities is observed. In this state  the nonthermal tail and blackbody    components become comparable in flux.    The power-law component has a photon index of $2.5$ and does not show evidence of a cutoff even out to a few hundred keV.
\item[\tt Quiescent/off state] In the last state the object appears  to be \lq off\rq,   with a flux level much lower than in the other states and a photon   index softer than in the low/hard state. \end{description}
\cite{R02}, and again \cite{RMc06} took another approach. They defined  a set of three \lq fundamental\rq~ states (using more descriptive names, not based on the X-ray luminosity) and the transitions between them. These three states are   the Low-Hard, Steep Power Law (Very High) and thermal (High-Soft) states, respectively LHS, SPL and thermal states hereafter. Other states are simply transitions between these three.  XTE J$1550$-$564$ is a good example of this classification. \\
Given the strong differences between the behaviors of the known microquasars, the differences between these classifications can arise simply from the sources that receive more attention in each work. It is however interesting to note already the strong link between these three classes and the temporal behavior of the sources: the thermal state does not show QPOs and the SPL has both  low-frequency and high-frequency QPOs (hereafter LFQPO and HFQPO).  
\refe{For the low-hard state the association is not as straightforward. 
This state often exhibits a LFQPO whose frequency has been shown to be correlated with various properties of the accretion disk, and therefore
ultimately to the accretion rate itself.}
\refee{One must note, however, that there are sources where the LFQPO is not detected in the LHS. 
There are several possibilities for this absence. Either the QPO is truly absent,
or the  amplitude of the QPO is too weak to be detected. The latter possibility can have various origin again. 
The low luminosity of the source may not permit to detect it, or the QPO frequency might simply fall out of the frequency range.
Another explanation is that the underlying instability mechanism giving rise to it is itself weak or fails to generate
a strong observable modulation of the X-ray emission. This can be the case if,  e.g., the physical properties of the disk and
its corona, or of its inclination (one must remember for instance that a simple rotating hot spot would not cause a QPO, even in a strongly inclined source)
 make the detection difficult, or the QPO falls out of the frequency range analyzed, or the QPO is truly absent. 
In a weak source such as XTE J$1752$-$322$, 
the LFQPO seems absent from the LHS, but appears weakly when observations are summed  \citep{MD10}. In a  stronger source such as 
XTE J$1550$-$564$ \citep{R03}, LFQPOs appear early in the initial Low-Hard state and immediately after the transition back to the Low-Hard state.
Given the very frequent observation  of LFQPOs in the LHS in other sources (H$1743$-$322$, e.g. Prat et al. 2009;
XTE J$1118$+$480$, Revnivtsev et al. 2000; XTE J$1908$+$094$, Gogus et al. 2004; $4$U $1630$-$474$ Dieters et al 2000)
 we consider it as a property, which may or not grow to be observable, of that state.}
\newline

\refee{The original three state description has been refined further using from the start}
X-ray timing properties, especially the type of Low-frequency Quasi-Periodic Oscillation, and  observations at other wavelengths
 such as the presence of a jet seen in radio, \refee{to separate the different states}. 
 \cite{HB05} introduced a 4-state classification in the continuity of the previous ones, but with a stronger link 
 to the QPOs observed, and in particular to the identification of three different flavors (A, B and C) of the LFQPO \citep{R02}.  
\refe{The High-Soft state is in the direct continuity of the previous classification while the Low-Hard state is defined similarly with the exception of the
observation where a type C LFQPO is in the observation. It mainly correspond the early part of outbursts.
 The two other states of their classification are the Hard Intermediate State,  in which a C-type LFQPO  is detected, while the Soft Intermediate state is characterized by either a B- or A-type LFQPO  \refee{\citep{HB05}}. The differences between those LFQPOs include the evolution of both frequency and amplitude, the harmonic content and the noise components such as the Band Limited Noise.}
 This classification is coherent with an evolution  from Low/Hard to  Hard Intermediate, Soft Intermediate, High Soft, Hard Intermediate and back to 
 Quiescent/Low-Hard State. 
 \refe{However, even this more refined classification does not encompass all the observed behavior; for example the case  of HFQPOs without
 LFQPOs (as seen in  \GRS in the B state) does not fit in any class.}\\
 
A different classification was defined by \cite{B00} for the special case of the  most spectacular and active microquasar, GRS $1915$+$105$. They defined $3$ fundamental spectral states, labeled A, B and C (not related to the A, B and C LFQPO types), which 
combine into $12$ reproducible classes of variability ($\alpha, \beta,$ ...). The three  states, defined in a color-color diagram, are again characterized 
by different contributions from the disk (thermal component of the X-ray) and its corona (non-thermal component at higher energy).  
Each of the 12 variability classes is either composed of only one state $A$ (class $\phi$), C ($\chi$) or  repetitive cycles of various length of BAB ($\delta, \gamma$), CAC ($\theta$) or CAB ($\alpha, \nu, \rho, \kappa, \lambda, \beta,\mu$).  This classification has proven useful,  but is limited to that particular object. The main difference between \GRS and the other microquasars, besides the fact that it has remained active since it was first observed,  is actually the occurrence of these cycles on short (up to a few tens of minutes) timescales, and their repeatability over the years.\\
There is no one-to-one connection between this spectral classification and the previous ones, partly because  \GRS tends to stay at a much higher flux level than other objects \refee{\citep[see {\em e.g.}][]{FB04}}. However, considering the short-term temporal behavior of the three basic spectral states, 
 a one-to-one association does  seem to exist: indeed, the A state of  \cite{B00} does not show any quasi-periodic oscillation,  while C always exhibits the Low-Frequency one. The B state seems to always be present when HFQPO are detected alone ({\em i.e.} without a LFQPO) though no extensive study has been published yet (see for example Morgan {\em et al.}, 1997). \\

These classifications are based on observations and their differences point to the differences between the sources used as reference. 
The SPL, which  might actually be defined by the joint observation of LF and HFQPOs, was often seen in \XTE  but  was not,
until recently, observed in  \GRS and therefore was absent from the  classification of   \cite{B00}. As we have already mentioned, the same can be 
said about the $B$ state of \cite{B00}, during which \GRS seems to exhibit a HFQPO alone. This has not yet been observed in \XTE and is therefore 
absent from the classification of \cite{R02} and \cite{RMc06}. \\

\refe{Here we take another approach. We decided to look at the different behaviors of microquasars from the perspective of 
the disk instabilities happening in them. We are focusing here  on instabilities that have clear observational signatures, 
namely the ones that could explain the QPOs (HF and LF).  We are making the assumption that the mechanisms (instabilities) at the origin of the QPOs 
become, when they happen, the dominant effect in the disk. 
\refee{Also, these instabilities will be impacting the disk even before the observable "QPO" can be detected with our present capabilities 
 (see for example the numerical work of Caunt \& Tagger, 2001, for the evolution of a disk with the Accretion-Ejection Instability).}
The fact that, in all the existing classifications, the different states could always be linked with a 
change in the QPO-content tend to agree with that assumption. 
Our goal is to classify the behavior using  what occurs in the disk. 
We therefore start from the different instabilities thought to be active in the accretion disk and look at the different observables  
they would lead to. Using this to create a classification allows it to go beyond the \lq not yet observed\rq~arguments as it is not based on any set of observations but on more general principles.} 

The first step in this direction was made with the magnetic flood scenario 
\citep[MFS:][]{TVRP04} which was presented as  a possible explanation for the $30$ minute cycle (class $\beta$) of  GRS $1915$+$105$. 
This scenario started from the tentative association of the LFQPO with the  Accretion-Ejection Instability \citep[AEI:][]{TP99, V02,  RV02}, 
and proceeded by a comparison between the observed properties of the source in the various states of  \cite{B00} and the physical requirements for 
and expected consequences  of the development of the AEI. 
The second step \citep{TV06} was to  propose the Rossby-Wave Instability (RWI) as a possible mechanism for the High-Frequency QPO in microquasars. 
In these works the properties of the disk expected to {\em result from} these instabilities were part of the argument for their identification with the QPOs.\\
In the present work we continue this effort by undertaking a classification that starts from the physics of the different instabilities that can exist in disks.
\refeee{Throughout this paper we will use one instability for the LFQPOs, without entering into their different types which will be treated in more detail in a forthcoming publication \citep{V10}, and one instability for the HFQPOs.}
 Comparing the conditions required for them to appear, and their expected consequences on the disk, we will then turn to the observations and seek the correspondence that can exist. \refe{As this is based on the instabilities at the origin of the QPOs, we will focus first on them and not seek any one-to-one
association with the observation-based classifications. } 
\\ 
In section 2 we will  thus briefly  review the three states of   \cite{B00}, and the three MHD instabilities that have been predicted in magnetized disks. We will then discuss how these instabilities can be accommodated within the Magnetic Flood Scenario and the three spectral states it involves. In the following section we will then show that the occurrence of the three instabilities can actually give \refe{{\bf four instability-defined states}}. In section 4 we will thus concentrate on the fourth state: we will first show it to occur in MHD disk simulations, and find confirmation in observations \refe{of several sources and not just \GRS from which we started}.  \refe{In section 5 we will look a this tentative classification through the instabilities occurring in the disk and see how it can be 
seen as a model independent classification just based on physical processes without naming the instabilities. This will allow us to focus on what this 
type of physics-based classification can give us compared with the phenomenological one.}
\section{Magnetic Flood: a three instabilities scenario}
\subsection{The three states of \cite{B00}}
\begin{figure}[htbp]
\centering
\resizebox{\hsize}{!}{\includegraphics{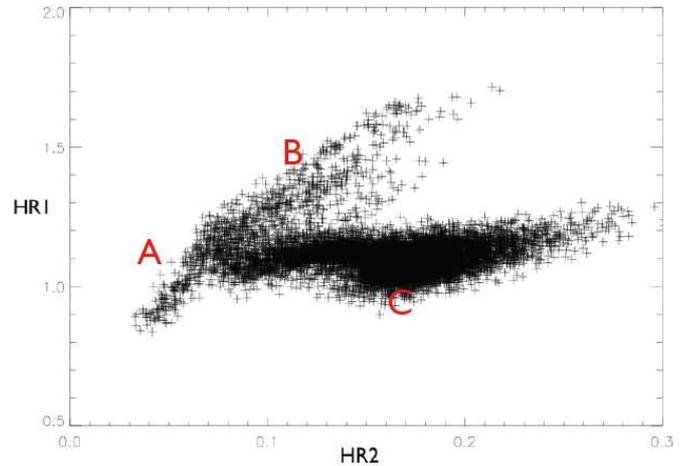}}   
\caption{three states of Belloni shown on the color-color diagram of the class $\beta$ of \GRS.}
\label{fig:abc}
\end{figure}

  \cite{B00} classified the behavior of the
microquasar GRS $1915$+$105$ in three fundamental spectral states labelled A, B and C, based on their position in a color-color diagram (see figure Fig.\ref{fig:abc}). \refe{With the hardness ratio defined as 
HR1 = B/A and HR2 = C/A where A corresponds to the channel $0-13$ ($2-5$ keV), B to $14-35$ ($5-13$keV) and C to $36-255$ ($13-60$keV).}\\

Their definitions of the three states, using color ratios HR1 and HR2, are as follows: 
\begin{itemize}
\item {\sf  state A:} This state is at low count rate, low HR1 and HR2 ratios, in the lower left corner of the color-color diagram. In this state the disk flux represents a substantial contribution to the total flux. There is no quasi-periodic variability and the X-ray emission  is mostly from  the thermal  disk emission.\\ 
In comparison with other classifications, this is closest to the High/Soft or Thermal state since there is a dominant  disk and no quasi-periodic variability.\\
\item{\sf  state C:} State C has a low count rate, low HR1 and variable HR2 depending on the length of the event. The power-law flux dominates with a little or no visible disk flux contribution. \\
This state is known to exhibit a LF-QPO and  Band-Limited Noise (BLN). \\ 
\refe{In comparison with other classifications, this is closest to the Low/Hard state}\\

\item{\sf  state B:} This state is located above state A in the CC with a high HR1. This state also has a  high count rate. Substantial red-noise variability is visible on time scales longer than 1s.
The disk is visible and hotter than in state A, and spectral fits indicate that it is close to the last stable orbit. High-Frequency
QPOs  alone are observed in this state.\\
\refe{There seems to exist no equivalent to this state in the other classifications.}\\
\end{itemize}    
\subsection{The three instabilities at the origin of the three states}
\label{sec:threeI}
Three instabilities have been proposed to exist in magnetized accretion disks. They differ by their physics, by the conditions for their existence and by their expected consequences:\\

{\sf Magneto-Rotationnal Instability (MRI)} \\  The MRI is a local instability that develops in a weakly magnetized disk  (below equipartition with gas pressure, \ie~a plasma $\beta = 8\pi p/B^2 > 1$),   when 
\begin{eqnarray}
\frac{\partial \Omega}{\partial r} < 0
\end{eqnarray}   
which is always true for (pseudo-)Keplerian disks.  
Numerical simulations have confirmed that it results in small-scale turbulence, and it is believed to be responsible for accretion in general. However no simulation of the MRI has yet shown a QPO of any kind. We thus assume that the corresponding variability, lacking any strong feature, is the one best adapted to correspond  to a power-law temporal spectrum \citep{BH02}. \\

{\sf Accretion-Ejection Instability (AEI)} \\

The AEI \citep{TP99} is a global instability occurring in disks threaded by a poloidal magnetic field close to the equipartition, namely when the plasma $\beta \sim 1$. These conditions are the ones found necessary in MHD models of jets \citep[see {\it e.g.}][]{CF00}\\

The AEI belongs to the same family as the spiral instability of self-gravitating disk galaxies, and the Papaloizou-Pringle Instability.  
In the language of diskoseismology \citep{W99, K01} these could be described as unstable p-modes, while the RWI discussed below would correspond to the g-mode.
The AEI is characterized   by a spiral structure that develops in the inner region of the disk.  
At the corotation radius, where the gas and the spiral wave rotate at the same angular velocity, the spiral excites a Rossby vortex to which it transfers the energy and angular momentum it extracts from the disk.  \\ %
Furthermore, in the presence of a low density corona, the Rossby vortex twists the footpoints of the magnetic field lines, generating an Alfv\'en wave propagating to the corona where it might provide the source for a wind or a jet \citep{VT02}.  This instability is stronger when the plasma $\beta$ is of the order of one and requires
\begin{eqnarray}
\frac{\partial}{\partial r} {\mathcal L}_{B} > 0
\end{eqnarray}
where 
\begin{eqnarray}
{\mathcal L}_{B}=\frac{\kappa^2 \Sigma}{2\Omega B^2} , 
\end{eqnarray}
$\Omega$ and $\kappa$ are the rotation and epicyclic frequencies (in a Keplerian  disk $\Omega = \kappa$), $\Sigma$ is the surface density and $B$ is the equilibrium magnetic field. This criterion is obeyed in disks with \lq reasonable\rq~density and magnetic field profiles, whereas for the Papaloizou-Pringle Instability in an unmagnetized disk the criterion becomes
\begin{eqnarray}
\frac{\partial}{\partial r} {\mathcal L}_{p} > 0
\end{eqnarray}
where 
\begin{eqnarray}
{\mathcal L}_{p}=\frac{\kappa^2}{2\Omega \Sigma}  
\end{eqnarray}
which would require a very steep density gradient to be destabilizing.\\
We have presented the AEI as a good candidate for the low-frequency 
quasi-periodic oscillation \citep{RV02, V02, M09} and discussed \citep{VT02} how the correlation observed between this QPO and coronal activity could be explained in relation with the emission of Alfv\'en waves. 
\refee{In a forthcoming publication \citep{V10} we will also show how the three types of LFQPOs can be explained by the AEI within the framework presented here.}\\

{\sf  Rossby Wave Instability (RWI)} \\

The RWI can develop in unmagnetized as well as magnetized disks, requiring respectively an extremum of ${\mathcal L}_{P}$ or ${\mathcal L}_{B}$. Such an extremum can be due to an extremum of density \citep{LOV99, TM06} or to relativistic effects near the Last Stable Orbit (LSO) in the accretion disk of a microquasar \citep{TV06}: indeed the LSO is defined by a vanishing epicyclic frequency $\kappa$, whereas further out in the disk $\kappa$ is close to $\Omega$. Thus $\kappa^{2}/2\Omega$, and consequently ${\mathcal L}_{P}$ and ${\mathcal L}_{B}$, have extrema near the LSO. This extremum of $\kappa$ occurs at $r\simeq 1.3 \  r_{LSO}$, so that the RWI can occur whenever the inner radius of the disk is within that radius.\\
As with the AEI, the RWI excites Alfv\'en waves and can be expected to energize the corona if the disk is threaded by a poloidal magnetic field, though the field is not necessary for the instability.
We have proposed the RWI as a possible explanation for the observed HFQPOs in microquasars \citep{TV06}. 
\subsection{Magnetic Flood Scenario} 
The Magnetic Flood Scenario (Tagger {\em et al.} 2004)  was first introduced to explain the $\beta$ class of the classification of  \citep{B00} (also known as 
 the $30$-minutes cycle). The MFS explains the repetitive X-ray behavior as a limit cycle determined by
the advection of poloidal magnetic flux to the inner disk and its destruction via magnetic reconnection (which can lead to relativistic ejection) with the
magnetic flux trapped close to the source.\\
The MFS starts with the identification of the Low Frequency Quasi-Periodic Oscillation with the AEI. It then proceeds by assuming that the onset of the
\refe{state C of the  Belloni {\em et al.} 2000 classification (possibly identified with the low-hard state in other classifications 
or the hard intermediate state in the classification by  Homan \& Belloni,  2005.)}, 
which also corresponds to the onset of the QPO, is triggered when the disk magnetization becomes large enough (of the order of equipartition with the gas pressure, so that we might call this a \lq fully magnetized\rq~disk) for the AEI to become unstable.\\ 
We also noted that state A was the one closest to what is expected from an $\alpha$-disk dominated by viscous-like transport of angular momentum due to small-scale turbulence. We thus proposed to associate state A with the presence of the Magneto-Rotational   Instability (MRI). Finally, we proposed recently  the Rossby-Wave Instability (RWI) as a possible explanation for   the HFQPO observed in microquasars \citep{TV06}. 
An interpretation of the B state of  \cite{B00} was thus called for, in which this state would be dominated by the RWI when the disk inner radius is 
close to the last stable orbit. \\
As a consequence we associated the three fundamental states of  \cite{B00} 
with three distinct instabilities, based on their properties and variabilities.

\section{Three Instabilities but four states}
Let us now discuss the instability criteria, for the three instabilities of section \ref{sec:threeI}, in terms of two parameters of the inner disk region: the location of its inner radius, measured by a parameter $\xi = r_{int}/r_{LSO}$, and the magnetization of its inner region, measured by $\beta=8\pi p/B^{2} $: 
\begin{description}
\item{- the MRI} requires a weakly magnetized disk, \ie $\beta>1$. It does not depend on $\xi$, and it is not believed to cause QPOs.
\item{- the RWI} requires an inner edge close to the LSO, \ie~$\xi<\xi_{ext}$ where $\xi_{ext}$ is the position of the extremum of either ${\cal L}_B$ or ${\cal L}_P$. 
It is stronger if $\beta$ is of the order of unity, but can exist even in an unmagnetized disk. It has been proposed to explain the HFQPO.
\item{- the AEI} requires a positive gradient of the quantity ${\mathcal L}_{B}$, which as discussed in section \ref{sec:threeI} is obtained with \lq reasonable\rq~assumptions on the radial profiles of $\Sigma$ and $B$. It also requires a magnetic field of the order of equipartition, \ie $\beta\sim 1$. Its existence does not depend on $\xi$, but its frequency is a fraction of the Keplerian rotation frequency at $r_{int}$, and thus a function of $\xi$.  It has been proposed to explain the LFQPO.
\end{description}
These criteria are not mutually exclusive. In particular, if the inner disk edge is close enough to the Last Stable Orbit (so that the RWI exists) and the magnetic field is sufficient, the AEI can exist further out in the disk, with a lower frequency. \\
Thus with the two parameters $\beta$ and $\xi$ we can map four cases, defining  four states of variability:
\begin{description}   
\item[$\beta >1$, $\xi_{int}  > \xi_{ext}$:]  The disk is weakly magnetized and is not at the  Last Stable Orbit \refee{even if it is really close to it}. 
The MRI must dominate, leading to a turbulent disk with no QPO. This state is similar to the High/Soft, Thermal or A state in previous classifications 
\refee{ as it does not display any QPOs}.  
 \item[$\beta >1$, $\xi_{int}  < \xi_{ext}$:] The disk is weakly magnetized and its inner edge is close to the Last Stable Orbit. The RWI is present at the 
 inner edge, while the MRI acts further out.  Observationally this should appear as a warm disk with a HFQPO and no LFQPO.
\item[$\beta \sim 1$, $\xi_{int} > \xi_{ext}$:] The disk is \lq fully magnetized\rq, \ie the field is of the order of equipartition
with the gas pressure, and the inner edge does not approach the Last Stable Orbit. 
The AEI causes a LFQPO so that the inner region of the disk cools down while the corona becomes active. 
\item[$\beta \sim 1$, $\xi_{int}< \xi_{ext}$:] The disk is fully magnetized and the inner edge is close to the Last Stable Orbit. Both the AEI and the magnetized version of the RWI are present, producing both a LF- and HF-QPO. The frequency of the LFQPO varies little since it is a fraction of the rotation frequency at the inner disk edge, which cannot change much. The interaction between these two modes can be presumed to affect their characteristics and their effect on the disk (see Varniere {\em et al.},  2010 \refe{for a comparison of those effects with the different types of LFQPO}).
\end{description}
Thus from the two parameters $\beta$ and $\xi$ we have defined four regions of parameter space, leading to four types of behavior. 
The three states that were discussed in the Magnetic Flood Scenario are easily associated with the states $A$, $B$ and $C$ of  \cite{B00}. 
\refe{Here, we arrived to the conclusion that a fourth state is required to describe all  possible behaviors. In this state} 
the AEI and the RWI are active, and thus there are both the LF- and HF-QPO.

\refe{It is also interesting to note that, while this separation in four state started from the QPOs alone, the presence of the Band-Limited Noise seems
to correlate with only one state, namely the AEI/LFQPO-only state.}

\section{The State \{AEI + MRI\}: $\beta \sim 1$, $r_{int}/r_{LSO} \sim 1$}
%
\refe{That fourth state is defined in the parameter space by the co-existence of two instabilities which lead to the observable that are
the HFQPO and LFQPO. In order to validate this, we need to show that a state with these observables is actually observed  {\bf and} that these
two instabilities can actually co-exist.}

\subsection{Observations of a state \refe{with both HF and LF QPOs}}
\refe{The MFS, based on observations of \GRS, is at the origin of the association of the three instabilities with the three A,B and C states. 
The fact that the fourth state (the one with both LF and HF QPO) was recently observed in this source  \citet{B06} is important as it gives strength and confidence in our tentative classification.}

%
On the other hand, while this fourth prescription had already been observed in other sources such as \XTE (so 
 this state is already part of the standard three/five-states description),  the HFQPO-only state has so far never been seen in any of those other objects.
\refe{There are a couple of
possibilities for why that state has not been observed yet: 1) the HFQPO is too weak to be detected or 
2) those sources are able to meet the criteria for the RWI only when they also meet the criteria for the AEI.
 In either case, it would be interesting to see what is singling out \GRS
 to either have its HFQPO alone strong enough to be detected or being able to meet the criteria for one of the HFQPO and not for the LFQPO at the same
  time. This would require a more extensive study
 of the observation and is beyond the scope of this paper.}
\subsection{Numerical Evidence of the \{AEI + RWI\} state} 
Since we have discussed the importance of observing the LF- and HF-QPO together, we now proceed to show that this can occur in 
numerical simulations too. This is numerically demanding since it requires coping with different time scales and different instability 
criteria, and we will thus stick to a proof-of-principle simulation, choosing parameters that make it easier rather than claiming realism.

\subsubsection{Numerical Setup}  
As in \cite{TV06} we use the 2.5D MHD code first introduced by \cite{C01} to study the non-linear evolution of the AEI. The code, based on a Zeus-type scheme, is similar to those  used for simulations of galactic spiral structures before full 3D simulations were possible; it describes an infinitely thin disk in vacuum. The magnetic field can be described by a magnetic potential outside the disk, related to the field in the disk by a Poisson equation similar to that for self-gravity. The code uses  cylindrical coordinates with a logarithmic radial grid. This allows high precision in the inner region of the disk where the instabilities develop, and a large  dynamic range in $r$ (typically $50$ inner radii) to avoid boundary condition issues.  
The code also  implements the FARGO scheme (Masset, 2000) which enhances execution speed by eliminating the Keplerian speed from the Courant condition. Finally the code mimics the relativistic rotation curve (needed for the RWI) with a pseudo-newtonian potential \citep{PW80}:
\begin{eqnarray}
\Phi = -\frac{GM}{r-r_s} \ \ \ \ \ \ \ \ \ \ \mbox{with} \ \ \ \ \ \    r_s = 2 GM/c^2 \nonumber
\end{eqnarray}
 \\
 
 This gives us 
 $\kappa_R = \Omega_R \sqrt{(r-3r_s)/(r-r_s)}$,  which has a maximun close to 
 the last stable orbit.\\ 
 
The conditions to obtain both instabilities are:\\
- a disk near equipartition between the gas pressure and the magnetic field, namely $\beta = 8\pi p/B^2 \sim 1$ \\
-  with its  inner edge  near  the last stable orbit, $\xi_{int} \approx 1$.  \\
- an extremum of ${\cal L}_B=\kappa^2/(2\Omega) \Sigma/B^2$ for the RWI\\
- and a positive derivative for ${\cal L}_B$ for the AEI\\

By using a pseudo-newtonian potential we already have an extremum of ${\cal L}_B$  close to the last stable orbit. We need 
to ensure that it also has a positive derivative further away in the disk so that the AEI can develop. 

We thus seek $(\kappa^2/(2\Omega)) \Sigma/B^2$ to have a positive gradient away from the last stable orbit.  For this we 
can use the Keplerian form of ${\cal L}_B^K=(\Omega_K/2)  \Sigma/B^2$. Away from the last stable orbit ${\cal L}_B^K \simeq{\cal L}_B$.
We will impose ${\cal L}_B^K \propto \xi^{0.1}$ and we will use it to fix the radial profile of B for a given density profile. 
This will ensure that we have the AEI developing in the disk. 

We then need to define the density $\Sigma$. Since we want a profile that is not too steep at the inner edge of the simulation, we take
\begin{eqnarray}
\Sigma\  &=& 2 \Sigma_o \left( \frac{1}{\sqrt{\xi}} - \frac{1}{2\xi}\right) \nonumber
\end{eqnarray}  

Finally we use the definition of ${\cal L}_B^K$ to set the magnetic field: 
\begin{eqnarray}
B    \  &=&  \sqrt{\frac{\Sigma}{{\cal L}_B^K \Omega_K}}\nonumber  
\end{eqnarray} 
  This setup will ensure that we have both the RWI and the AEI appearing in the inner region of the disk.
 
\subsubsection{Results}
\begin{figure}[htbp]
\centering
\resizebox{\hsize}{!}{\includegraphics{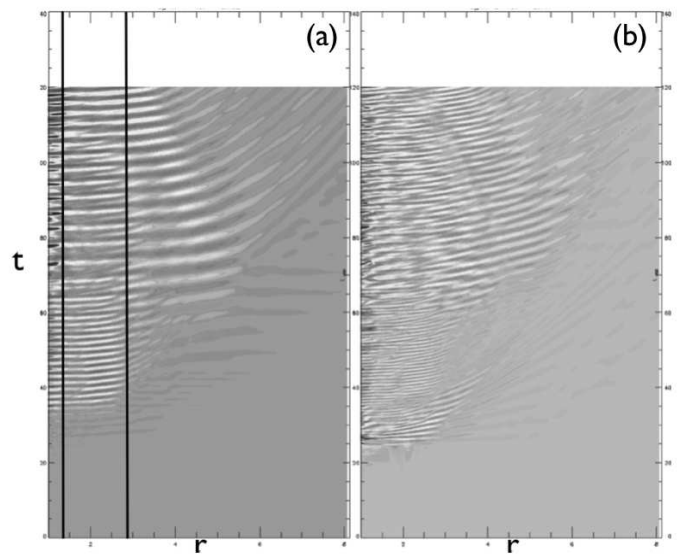}}  
 \caption{Contour plot of modes $m=1$ (a) and $m=2$ (b) of the radial velocity as a function of radius and time. 
 }
\label{fig:2_instabilite1}
\end{figure}

  Figure \ref{fig:2_instabilite1}  shows a contour plot  of the amplitude of $m=1$ and $m=2$ perturbations in radial velocity, as a function of radius and time.  These plots are a convenient way to study waves as they allow us to identify their propagation, \ie oblique features showing a traveling wave while horizontal ones indicate a standing pattern. The AEI appears as a standing pattern within its corotation radius, emitting a wave traveling outward beyond it. The RWI is a standing pattern trapped in the extremum of $\mathcal L_{B}$ and a traveling wave elsewhere.\\
On Fig.\ref{fig:2_instabilite1} we see at first the $m=1$ and $m=2$ modes\footnote{We choose not the show all the modes but to focus of the first two for simplicity. } of the RWI close to the relativistic maximum of  $\cal L_{B}$ (which occurs in our simulation around $1.3\ r_{LSO}$). 
Further away, we also see the $m=1$ and $m=2$ modes of the AEI.      

Having both instabilities present tends to make the spectrum messier compared to having them separately,  as they are not fully radially separated. 
In a companion paper \citep{V10} we will focus on how the instabilities are modified in that case and how it could explain some of the features of the different types of LFQPOs.
We insist that, given the crudeness of the model, this numerical simulation can only be considered as a proof-of-principle. 
In particular the value of $m$ that is most unstable, and thus dominant, depends on the profiles we have used, but also on the quality of 
the model. Indeed in separate work on unmagnetized disks \citep{M10} we find that $3$D  simulations can give a result differing from the $2$D one.\\  

\section{An \refe{Instability}-Based Classification}

\subsection{A model Independent look at the classification}
\begin{figure}[htbp]
\centering
\resizebox{\hsize}{!}{\includegraphics{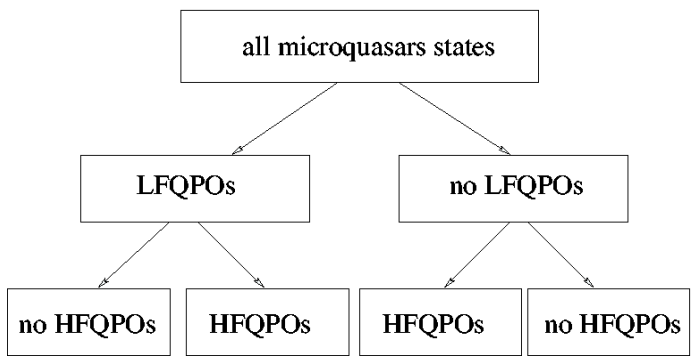}}  
\caption{{\small Model independant view of our classification in a tree form.}  }
\label{fig:classification1}
\end{figure}

Although we have discussed it in view of definite QPO models, the classification we propose is only based on the 
{\bf \refe{\refee{predicted} presence or absence} of LF and HF QPOs}. As such it is model-independent and can be compared with other classifications starting from the 
spectral properties. Implicitly this means that we consider the QPOs to be the {\em cause} of the spectral properties, assumed to depend on the 
transport and deposition in the disk or the corona of the accretion energy. Spectral-based classifications, on the other hand, 
simply describe the QPOs as properties of the spectral states. 
We find that these spectral properties agree well with what we can expect from our QPO models, but this may not be unique to these models \refe{and therefore
any model agreeing with the spectral properties and having the possibilities of both HF and LF QPOs would give the same four-states classification.} \\
This leads us to four states, defined by the presence or the absence of HF and/or LF QPOs. We can represent these four states as a tree diagram, shown in figure \ref{fig:classification1}. \\
This classification is actually close to the ones of  both \cite{B00} {\bf and} of \citep{R06, RMc06}.
The main difference with the classification of  \cite{B00}, based on GRS $1915$+$105$, is that we separate their C state in two 
(AEI-dominated state and AEI-RWI-dominated state),  depending on the high frequency variability. 
Since this classification was first proposed, we have now learned that HFQPOs occur in the C state as defined by \cite{B00} \citep[see][]{B06}, 
so that the AEI-MRI-dominated  state has been observed in GRS 1915+105.\\
Concerning the classification, mostly based on \XTE, in \citep{R06, RMc06}, the main difference is the
state with only HFQPOs which has not been observed yet and was not part of their classification. \refe{It will be interesting to look for reasons
why that state is observed in the case of \GRS and not for other objects.}
%
\refe{The major differences between
our classification and the one of Homan \& Belloni is that they do not have the HFQPO alone state} \refee{and that they separate the Low-Hard state and the Hard intermediate state based on multi-wavelength properties (nevertheless, they clearly state that their components are clearly related). }

\subsection{A method to find more HFQPOs?}

\refe{As one can see this instability-based classification is coherent with the other classifications based on a combination of spectral and timing behavior. 
This strengthens our hypothesis of a strong causal link between timing and spectral changes. This classification aims at creating a framework to 
better understand what drives the source behavior and is not to be used instead of a spectral classification, easier to use when dealing with observation.}
Indeed, having a classification based on the detection/non-detection of HF-LF QPOs has a limited interest
 \refe{when one is looking purely at classifying a state for reference} not only because HFQPOs are hard to detect but also they 
 require some processing of the data.
 
On the other hand, we can turn around that difficulty and use this classification \refe{as a tool} to probe for HFQPOs. 
Indeed, one can focus the search in the states that are clearly not-MRI or AEI-dominated. \refe{This is similar to the method \citet{R06} used to detect
HFQPO in H$1743$-$322$ following their discovery by \citet{H05}.}

In a second paper \citep{V10} we will focus  on the state where both the AEI and the RWI are present in the disk. 
\refe{In particular, we will study the impact of having both instabilities, more precisely the link between  HFQPO and LFQPO, and if the effect on the
LFQPO can be used to infer the presence of a weaker HFQPO.}

\subsection{\refe{When the instability criteria are met or not}}

\refe{As we showed in sect. 5.1  this model-independant classification in four states is only based on the presence or absence of the LF and HF QPO and not
on the actual instabilities at their origin. It is interesting to notice that most objects do not seem to exhibit one of the four states, namely the one with the HFQPO-only. Up to now,  \GRS is the only source known to exhibit all of the four states with some regularity. As this source was always singled out it is not a surprise to find in it behaviors that other sources do not exhibit, but with this classification we have a way to} \refe{ tackle the problem. 
Indeed, there are only 2 reasons why those HFQPOs are not detected in other sources:\\ 
1) the HFQPOs are present but are not detected because they are too weak. \\
2) the instability criteria for the instability at the origin of the HFQPO alone (whichever instability is the \lq right\rq one) is not fulfilled without the criteria for the LFQO being also fulfilled.\\
 In either case, it would be interesting to understand what}
 \refe{ is singling out \GRS to either have its HFQPO alone strong enough to be detected 
 or be able to meet the criteria  for a HFQPO without that for the LFQPO. }
 \refe{To answer that question would require an extensive study of the observation to decide in which case the other sources are
  and is beyond the scope of this paper.}


\section{Conclusion}

\refe{In this paper we have discussed several disk instabilities thought to occur in microquasars. From their instability criteria we
 can divide the observed parameter space in  four states characterized by their QPO content. 
This form a classification of all microquasars observations as shown on the tree diagram \ref{fig:classification1}.}
We find that this new classification re-unites previous  ones, which started from the spectral properties of the sources but, 
as a consequence, did not apply uniformly to all microquasars.\\
 Our discussion is guided by  the instabilities that may be at the
origin of the variability. Following previous works we have taken the Accretion-Ejection Instability to be at the origin of the LFQPO, and
the Rossby Wave Instability to be at the origin of the HFQPO, while the turbulent, thermal, disk state is assumed  
to be dominated by the Magneto-Rotationnal Instability. \\
Considering the conditions for these instabilities to develop we find, and confirm by numerical simulations as well as in recent observations,
 that  four states can be defined rather than the usual two, three or five states classifications used up to now, which can now be included in the one we present here. 
 The four states are defined by no QPO, only the LFQPO, only the HFQPO, and both LF- and HF-QPO, thus covering in a simple manner all the possible observations of QPOs. Furthermore this classification can be used as a guide in searching for occurrences of the HF-QPO, which is often difficult to detect.

\begin{acknowledgements}
The author thanks the anonymous referee that helped  clarify the paper to this final form.
This work has been financially supported by the GdR PCHE in France.
JR acknowlegdes partial funding from the European Community's
Seventh Framework Programme (FP7/2007-2013) under
grant agreement number ITN 215212 "Black Hole Universe.
\end{acknowledgements}

\end{document}